# Generation of Optical Chirality Patterns with Plane Waves, Evanescent Waves and Surface Plasmon Waves

Jiwei Zhang,[1,*] Shiang-Yu Huang,[1] Zhan-Hong Lin[1] and Jer-Shing Huang[1,2,3,4,*]

[1]*Leibniz Institute of Photonic Technology, Albert-Einstein Straße 9, 07745 Jena, Germany*
[2]*Abbe Center of Photonics, Friedrich-Schiller University Jena, Jena, Germany*
[3]*Research Center for Applied Sciences, Academia Sinica, 128 Sec. 2, Academia Road, Nankang District, 11529 Taipei, Taiwan*
[4]*Department of Electrophysics, National Chiao Tung University, 1001 University Road, 30010 Hsinchu, Taiwan*

e-mail address: jiwei.zhang@leibniz-ipht.de; jer-shing.huang@leibniz-ipht.de

## Abstract

We systematically investigate the generation of optical chirality patterns by applying the superposition of two waves in three scenarios, namely plane waves in free space, evanescent waves of totally reflected light at dielectric interface and propagating surface plasmon waves on a metallic surface. In each scenario, the general analytical solution of the optical chirality pattern is derived for different polarization states and propagating directions of the two waves. The analytical solutions are verified by numerical simulations. Spatially structured optical chirality patterns can be generated in all scenarios if the incident polarization states and propagation directions are correctly chosen. Optical chirality enhancement can be obtained from the constructive interference of free-space circularly polarized light or enhanced evanescent waves of totally reflected light. Surface plasmon waves do not provide enhanced optical chirality unless the near-field intensity enhancement is sufficiently high. The structured optical chirality patterns may find applications in chirality sorting, chiral imaging and circular dichroism spectroscopy.

# TOC Graphic

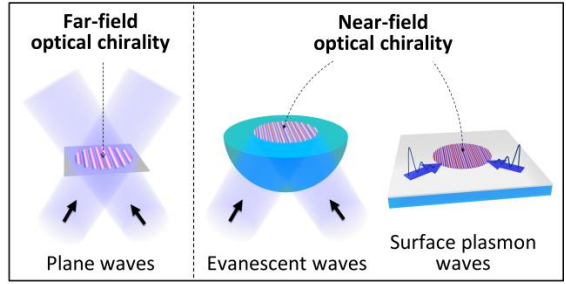

# Introduction

A chiral object can exist in two different forms that are non-superposable mirror images of each other. This asymmetry, called chirality, can be found at scales ranging from a tiny molecule to a giant galaxy. Similarly, an optical field can also be chiral. For example, circularly polarized light (CPL) and elliptically polarized light (EPL) are chiral and possess different handedness. The chirality of an optical field can be quantified by a conservative quantity called "optical chirality (OC)"[1,2]

$$C \equiv \frac{\varepsilon_0}{2} \mathbf{E} \cdot \nabla \times \mathbf{E} + \frac{1}{2\mu_0} \mathbf{B} \cdot \nabla \times \mathbf{B} = -\frac{\varepsilon_0 \omega}{2} \mathrm{Im}(\widetilde{\mathbf{E}}^* \cdot \widetilde{\mathbf{B}}), \tag{1}$$

where $\varepsilon_0$ and $\mu_0$ are the permittivity and permeability of free space, respectively, and $\omega$ is the angular frequency of light. **E** and **B** are the real parts of the complex local electric $\widetilde{\mathbf{E}}$ and magnetic field $\widetilde{\mathbf{B}}$. This quantity has been linked to the characterization of dissymmetry of chiral molecules.[2] This has led to great research interest in OC engineering for the enhancement of chiral light-matter interaction. For instance, engineered "superchiral" fields[3-5] with larger OC than that of CPL can enhance the chiroptical response of chiral molecules. In this context, various OC engineering approaches using plasmonic nanostructures[6-23], dielectric nanostructures[24-28] and one-dimensional photonic crystals[29,30] have been proposed to obtain enhanced chiral optical fields.

Apart from enhancing OC, spatially structuring OC has been shown promising for chirality sorting[31-41], circular dichroism (CD) measurement[42] and super-resolution chiral imaging[43]. For example, an optical field can induce a time-averaged optical force on chiral objects with a reactive and a dissipative component, where the reactive one is proportional to the spatial gradient of the OC patterns.[31] Spatially modulated OC patterns with uniform intensity have also been applied to acquire CD spectrum within a single camera snapshot.[42] Another example is the recent theoretical demonstration of chiral structured illumination microscopy, where spatially structured OC patterns are employed to form moiré patterns with chiral samples and thereby obtain images of chiral domains at sub-wavelength resolution.[43] In addition, inhomogeneous OC can be used to imprint chiral patterns in materials, e.g., liquid crystal polymers, which may find applications in imaging the helicity of light.[44]

According to eq (1), a chiral optical field can be obtained only when its electric field has components which are parallel to and not in phase with its magnetic field. In the far-field region, linearly polarized light (LPL) has no OC, whereas CPL has non-zero and spatially uniform OC. Similar to the generation of light intensity patterns by the superposition of coherent waves, it's also possible to generate structured OC patterns using wave superposition.[34,44,45] In this work, we comprehensively investigate the generation of OC patterns by the superposition of (i) two plane waves in free space, (ii) the near field of two total internal reflection (TIR)-based evanescent waves (EWs) and (iii) the near field of two propagating surface plasmon waves (SPWs). Different polarization states and incident directions of the two waves are considered in the three scenarios. Analytical solutions of the produced OC patterns for each scenario have been derived and verified by numerical simulations. The results show that spatially structured OC patterns can be generated in both far field and near field. OC enhancement can be obtained in three conditions, namely the constructive interference of free-space CPL, enhanced near field of TIR-based EWs, and SPWs with sufficiently high near-field intensity enhancement. This work provides a guideline for generating OC patterns by the superposition of two waves.

## Results and Discussion

### Superposition of two plane waves

We first investigate the superposition of two plane waves propagating in free space. The wave vectors can be described as $\mathbf{k}_{1,2} = k_0(\sin\alpha_{1,2}\cos\phi_{1,2}\hat{\mathbf{x}} + \sin\alpha_{1,2}\sin\phi_{1,2}\hat{\mathbf{y}} + \cos\alpha_{1,2}\hat{\mathbf{z}})$, where $k_0$ is the wavenumber of light in vacuum, $\alpha_{1,2}$ is the incident angle, and $\phi_{1,2}$ is the orientation angle respect to the +$x$ axis, as depicted in Fig. 1. Here, the numbers "1" and "2" denote the first and the second wave, respectively. The superposition of the two plane waves creates an overlapping volume. In this work, we focus on the OC patterns on the *xy* plane with the largest cross section. In principle, the polarization state of a plane wave can be described by any arbitrary set of two orthogonal polarization components. For simplicity, we have chosen to use the *s*-polarized (*s*-pol.) and *p*-polarized (*p*-pol.) components relative to the incident plane to define the polarization state of each wave. The *s*-pol. component has a phase difference $\Delta\theta_{1,2}$ with respect to the *p*-pol. one. When $\Delta\theta_{1,2} = 0$ or $\pm\frac{\pi}{2}$, the plane wave represents LPL or CPL. Other arbitrary $\Delta\theta_{1,2}$ corresponds to EPL. The complex electric field of the *s*-pol. component for each plane wave is

$$\tilde{\mathbf{E}}_{1s,2s} = E_{1s,2s} e^{i\mathbf{k}_{1,2}\cdot\mathbf{r}+i(-\omega t+\varphi_{1,2})} e^{i\Delta\theta_{1,2}}(-\sin\phi_{1,2}\hat{\mathbf{x}} + \cos\phi_{1,2}\hat{\mathbf{y}}), \tag{2}$$

where $\mathbf{r} = x\hat{\mathbf{x}} + y\hat{\mathbf{y}} + z\hat{\mathbf{z}}$ is the spatial coordinate, $E_{1s,2s}$ is the amplitude and $\varphi_{1,2}$ the initial phase. Accordingly, the complex magnetic field of the *s*-pol. component can be expressed as

$$\tilde{\mathbf{B}}_{1s,2s} = \frac{k_0}{\omega} E_{1s,2s} e^{i\mathbf{k}_{1,2}\cdot\mathbf{r}+i(-\omega t+\varphi_{1,2})} e^{i\Delta\theta_{1,2}}(-\cos\alpha_{1,2}\cos\phi_{1,2}\hat{\mathbf{x}} - \cos\alpha_{1,2}\sin\phi_{1,2}\hat{\mathbf{y}} + \sin\alpha_{1,2}\hat{\mathbf{z}}). \tag{3}$$

For the *p*-pol. component, the complex magnetic and electric field are given by

$$\tilde{\mathbf{B}}_{1p,2p} = B_{1p,2p} e^{i\mathbf{k}_{1,2}\cdot\mathbf{r}+i(-\omega t+\varphi_{1,2})}(-\sin\phi_{1,2}\hat{\mathbf{x}} + \cos\phi_{1,2}\hat{\mathbf{y}}), \tag{4}$$

$$\tilde{\mathbf{E}}_{1p,2p} = -\frac{c^2 k_0}{\omega} B_{1p,2p} e^{i\mathbf{k}_{1,2}\cdot\mathbf{r}+i(-\omega t+\varphi_{1,2})}(-\cos\alpha_{1,2}\cos\phi_{1,2}\hat{\mathbf{x}} - \cos\alpha_{1,2}\sin\phi_{1,2}\hat{\mathbf{y}} + \sin\alpha_{1,2}\hat{\mathbf{z}}), \tag{5}$$

where $B_{1p,2p}$ is the magnetic field amplitude. Upon superposition, the total electric and magnetic field are $\tilde{\mathbf{E}} = \sum_j \tilde{\mathbf{E}}_j$ and $\tilde{\mathbf{B}} = \sum_j \tilde{\mathbf{B}}_j$ ($j = 1s, 2s, 1p, 2p$), respectively. Taking $k_0 = \frac{\omega}{c}$ and $B_j = \frac{E_j}{c}$, the OC distribution of the superimposed plane waves calculated by eq (1) is

$$\begin{aligned}C = -\frac{\varepsilon_0 k_0}{2} \operatorname{Im}\{ & E_{1s}E_{1p}e^{-i\Delta\theta_1} - E_{1s}E_{1p}e^{i\Delta\theta_1} + E_{2s}E_{2p}e^{-i\Delta\theta_2} - E_{2s}E_{2p}e^{i\Delta\theta_2} \\ & + e^{i\Phi}[\cos(\phi_1 - \phi_2)(E_{1s}E_{2p}e^{-i\Delta\theta_1} - \cos\alpha_1\cos\alpha_2 E_{1p}E_{2s}e^{i\Delta\theta_2}) \\ & + \sin(\phi_1 - \phi_2)(\cos\alpha_2 E_{1s}E_{2s}e^{-i(\Delta\theta_1-\Delta\theta_2)} + \cos\alpha_1 E_{1p}E_{2p}) - \sin\alpha_1\sin\alpha_2 E_{1p}E_{2s}e^{i\Delta\theta_2}] \\ & + e^{-i\Phi}[\cos(\phi_1 - \phi_2)(E_{1p}E_{2s}e^{-i\Delta\theta_2} - \cos\alpha_1\cos\alpha_2 E_{1s}E_{2p}e^{i\Delta\theta_1}) \\ & - \sin(\phi_1 - \phi_2)(\cos\alpha_1 E_{1s}E_{2s}e^{i(\Delta\theta_1-\Delta\theta_2)} + \cos\alpha_2 E_{1p}E_{2p}) - \sin\alpha_1\sin\alpha_2 E_{1s}E_{2p}e^{i\Delta\theta_1}]\}, \tag{6}\end{aligned}$$

where $\Phi = k_0[x(\sin\alpha_2\cos\phi_2 - \sin\alpha_1\cos\phi_1) + y(\sin\alpha_2\sin\phi_2 - \sin\alpha_1\sin\phi_1) + z(\cos\alpha_2 - \cos\alpha_1)] + \varphi_2 - \varphi_1$. From eq (6), one can get the analytical solution of OC for a specific case by setting the corresponding quantities $E_j$, $\alpha_{1,2}$, $\phi_{1,2}$, $\Delta\theta_{1,2}$ and $\varphi_{1,2}$. In this work, we focus on two representative configurations of the two propagating plane waves, i.e., the cross-propagation ($\phi_1 = 0$, $\phi_2 = \frac{\pi}{2}$) and the counter-propagation configuration ($\phi_1 = 0$, $\phi_2 = \pi$).

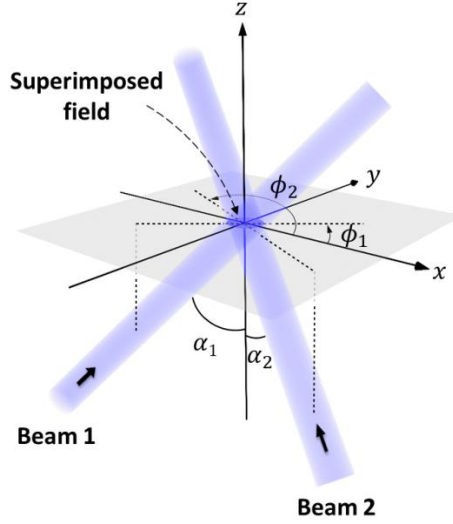

**Figure 1.** Schematic of two free-space plane waves at the same wavelength propagating along the *z*-axis at incident angle $\alpha_{1,2}$ and orientation angle $\phi_{1,2}$ respect to the +*x* axis.

To characterize the OC enhancement, we set the intensity of the superimposed field to be the same as that of free-space CPL. The normalized OC (noted as $\hat{C}$) is then obtained by calculating the ratio $\frac{C}{C_{\text{CPL}}}$, where $C_{\text{CPL}}$ is the magnitude of the OC of CPL. Here, we show the results of some commonly used cases where the two plane waves are LPL or CPL ($\Delta\theta_{1,2} = 0$ or $\pm\frac{\pi}{2}$) at the same incident angle ($\alpha_1 = \alpha_2 = \alpha$). The analytical solutions of the normalized OC for several exemplary cases are calculated and listed in the second and the third column of Table 1. The results show that chiral optical fields can be generated even with achiral LPL beams and the OC patterns are spatially structured, as indicated by the position-dependent term $\sin\Phi$ or $\cos\Phi$. On the other hand, using CPL beams does not guarantee non-zero OC. For example, the superimposed field is achiral ($\hat{C} = 0$) if two CPL beams with the opposite handedness are applied.

| Configuration | Cross-propagation ($\phi_1 = 0, \phi_2 = \frac{\pi}{2}$) | Counter-propagation ($\phi_1 = 0, \phi_2 = \pi$) |
|---|---|---|
| Plane waves | 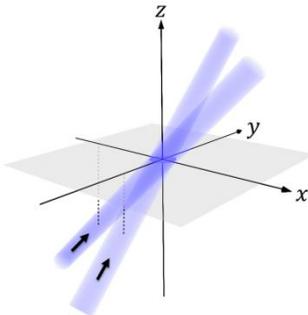 | 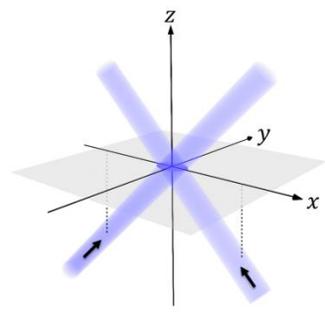 |
| | $\Phi = -k_0(x+y)\sin\alpha + \varphi_2 - \varphi_1$ | $\Phi = -2k_0 x\sin\alpha + \varphi_2 - \varphi_1$ |
| | $k_C = \sqrt{2} k_0 \sin\alpha$ | $k_C = 2k_0 \sin\alpha$ |
| $s$- and $p$-pol. light beam | $\hat{C} = -\frac{1}{2}\sin^2\alpha \sin\Phi$ | $\hat{C} = \cos^2\alpha \sin\Phi$ |
| two $p$- or $s$-pol. light beams | $\hat{C} = \cos\alpha \sin\Phi$ | $\hat{C} = 0$ |
| two CPL beams with same handedness (e.g., left-handed) | $\hat{C} = 1 + \cos\alpha \sin\Phi + \frac{1}{2}\sin^2\alpha \cos\Phi$ | $\hat{C} = 1 - \cos^2\alpha \cos\Phi$ |
| two CPL beams with opposite handedness | $\hat{C} = 0$ | $\hat{C} = 0$ |
| LPL (e.g., $p$-pol.) and CPL (e.g., left-handed) beam | $\hat{C} = \frac{1}{3}(2 + 2\cos\alpha \sin\Phi + \sin^2\alpha \cos\Phi)$ | $\hat{C} = \frac{2}{3}(1 - \cos^2\alpha \cos\Phi)$ |

**Table 1.** Analytical solutions of the normalized OC generated by the superposition of two free-space plane waves cross-propagating (second column) and counter-propagating (third column) at incident angles of $\alpha$.

Finite-difference time-domain method (FDTD Solutions, Lumerical) has been applied to numerically simulate the analytically predicted OC patterns in Table 1. The OC distribution of a left-handed CPL beam propagating in free space is also simulated in order to provide the value of $C_{\text{CPL}}$ for normalization. Details of the simulations are given in the Methods. The analytically calculated patterns are well reproduced in the FDTD simulations (Figs. 2(a)-(j)). With LPL beams, OC patterns with spatially alternating handedness can be generated (Figs. 2(a), (b) and (f)), except for the case of two counter-propagating $p$-pol. or $s$-pol. light beams (Fig. 2(g)). The exception comes from the fact that there is no electric field component parallel to the magnetic field component in the superimposed field. The amplitude of the OC patterns generated by the superposition of two LPL beams is always smaller than $C_{\text{CPL}}$. If two CPL beams with the same handedness are superimposed, structured and enhanced OC patterns with a single handedness can be generated (Figs. 2(c) and (h)). The enhancement in OC is due to the constructive interference of the electromagnetic fields. However, if two CPL beams possess the opposite handedness, the superimposed field would exhibit zero OC (Figs. 2(d) and (i)). Finally, the superposition of a LPL and a CPL beam can also lead to structured OC patterns possessing the handedness determined by the incident CPL beam (Figs. 2(e) and (j)). In this case, the OC enhancement is relatively lower compared to Figs. 2(c) and 2(h). For the cross-propagation and counter-propagation configuration listed in the second and the third column of Table 1, the spatial frequency of the structured OC patterns is $k_c = \sqrt{2}k_0\sin\alpha$ and $k_c = 2k_0\sin\alpha$, respectively. Basically, the spatial frequency increases with the increase of wavenumber $k_0$, incident angle $\alpha$ and orientation angle difference $|\phi_1 - \phi_2|$ of the two plane waves.

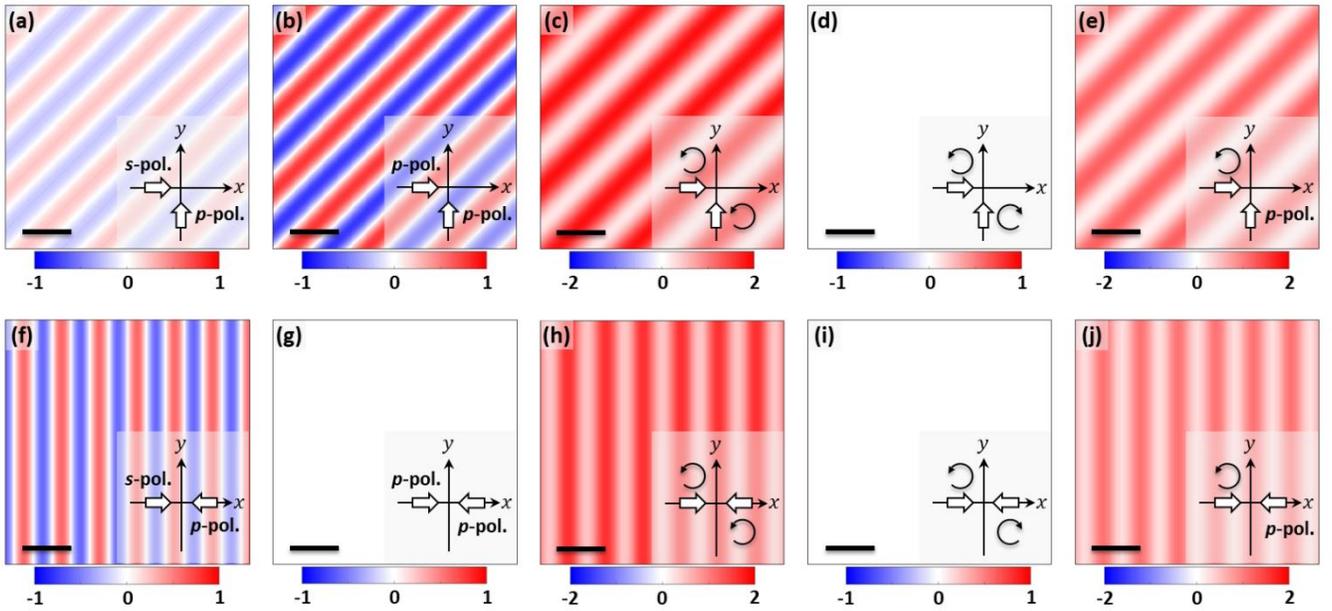

**Figure 2.** Numerically simulated normalized OC patterns generated by the superposition of (a,f) a *s*- and a *p*-pol. light beam; (b,g) two *p*- or *s*-pol. light beams; (c,h) two same-handed CPL beams (e.g., left-handed); (d,i) two opposite-handed CPL beams and (e,j) a *p*-pol. and a left-handed CPL beam. The two beams are (a-e) cross-propagating and (f-j) counter-propagating at an incident angle of $\alpha_1 = \alpha_2 = 41°$. The insets indicate the polarization states of the two beams and the projection of the incident directions on the *xy* plane. Scale bars: $2\pi/k_0$.

## Superposition of two evanescent waves

Previously, it has been demonstrated that two cross-propagating EWs with transverse electric modes in TIR can generate patterns of local circular polarization in the optical near field.[46] In this section, we derive the general solution of OC for the superposition of two EWs in TIR. As depicted in Fig. 3, two incident plane waves with wave vectors $\mathbf{k}_{1,2} = nk_0(\sin\alpha_{1,2}\cos\phi_{1,2}\hat{\mathbf{x}} + \sin\alpha_{1,2}\sin\phi_{1,2}\hat{\mathbf{y}} + \cos\alpha_{1,2}\hat{\mathbf{z}})$ are totally reflected at the prism-air interface when the incident angle $\alpha_{1,2}$ is larger than the critical angle of TIR $\alpha_c = \sin^{-1}(1/n)$, where $n$ is the refractive index of the prism. The superimposed near field is investigated within the overlapping area of the two EWs on the prism surface (*xy* plane, Fig. 3). Similar to the previous scenario, each incident plane wave is defined by the *s*- and *p*-pol. component with a phase difference of $\Delta\theta_{1,2}$. The incident angles are set to be the same, i.e., $\alpha_1 = \alpha_2 = \alpha$. Under these conditions, the complex electric and magnetic field of the EWs excited by the *s*- and *p*-pol. component of each incident light beam are

$$\tilde{\mathbf{E}}_{1s,2s} = t_s(\alpha)E_{1s,2s}e^{-\kappa(\alpha)z}e^{ink_0(x\sin\alpha\cos\phi_{1,2}+y\sin\alpha\sin\phi_{1,2})+i(-\omega t+\varphi_{1,2})}e^{i\Delta\theta_{1,2}}(-\sin\phi_{1,2}\hat{\mathbf{x}} + \cos\phi_{1,2}\hat{\mathbf{y}}), \quad (7)$$

$$\tilde{\mathbf{B}}_{1s,2s} = \frac{1}{i\omega}t_s(\alpha)E_{1s,2s}e^{-\kappa(\alpha)z}e^{ink_0(x\sin\alpha\cos\phi_{1,2}+y\sin\alpha\sin\phi_{1,2})+i(-\omega t+\varphi_{1,2})}e^{i\Delta\theta_{1,2}}$$
$$\times [\kappa(\alpha)\cos\phi_{1,2}\hat{\mathbf{x}} + \kappa(\alpha)\sin\phi_{1,2}\hat{\mathbf{y}} + ink_0\sin\alpha\hat{\mathbf{z}}]; \quad (8)$$

$$\tilde{\mathbf{B}}_{1p,2p} = t_p(\alpha)B_{1p,2p}e^{-\kappa(\alpha)z}e^{ink_0(x\sin\alpha\cos\phi_{1,2}+y\sin\alpha\sin\phi_{1,2})+i(-\omega t+\varphi_{1,2})}(-\sin\phi_{1,2}\hat{\mathbf{x}} + \cos\phi_{1,2}\hat{\mathbf{y}}), \quad (9)$$

$$\tilde{\mathbf{E}}_{1p,2p} = -\frac{c^2}{i\omega}t_p(\alpha)B_{1p,2p}e^{-\kappa(\alpha)z}e^{ink_0(x\sin\alpha\cos\phi_{1,2}+y\sin\alpha\sin\phi_{1,2})+i(-\omega t+\varphi_{1,2})}$$
$$\times [\kappa(\alpha)\cos\phi_{1,2}\hat{\mathbf{x}} + \kappa(\alpha)\sin\phi_{1,2}\hat{\mathbf{y}} + ink_0\sin\alpha\hat{\mathbf{z}}]. \quad (10)$$

Here, $\kappa(\alpha) = k_0\sqrt{(n\sin\alpha)^2 - 1}$ determines the penetration depth of the EW, $t_s(\alpha) = \frac{2n\cos\alpha}{n\cos\alpha+i\frac{\kappa(\alpha)}{k_0}}$ and $t_p(\alpha) = \frac{2n\cos\alpha}{\cos\alpha+in\frac{\kappa(\alpha)}{k_0}}$ is the complex transmission coefficient determined by Fresnel equations for the *s*- and *p*-pol. component, respectively.[47] $E_{1s,2s}$ is the electric field amplitude of each incident *s*-pol. component, $B_{1p,2p}$ is the magnetic field amplitude of each incident *p*-pol. component, and $\varphi_{1,2}$ is the initial phase of each incident light beam. By taking into account all the electric and magnetic field components from the two incident beams, the OC distribution of the superimposed EWs is

$$C = -\frac{\varepsilon_0 k_0 e^{-2\kappa(\alpha)z}}{2}\mathrm{Im}\{-t(\alpha)(2n^2\sin^2\alpha - 1)(E_{1s}E_{1p}e^{i\Delta\theta_1} + E_{2s}E_{2p}e^{i\Delta\theta_2})$$
$$+t^*(\alpha)(E_{1s}E_{1p}e^{-i\Delta\theta_1} + E_{2s}E_{2p}e^{-i\Delta\theta_2})$$
$$+t^*(\alpha)\cos(\phi_1 - \phi_2)(E_{1s}E_{2p}e^{-i\Delta\theta_1}e^{i\Phi} + E_{1p}E_{2s}e^{-i\Delta\theta_2}e^{-i\Phi})$$
$$-t(\alpha)[(n^2\sin^2\alpha - 1)\cos(\phi_1 - \phi_2) + n^2\sin^2\alpha](E_{1p}E_{2s}e^{i\Delta\theta_2}e^{i\Phi} + E_{1s}E_{2p}e^{i\Delta\theta_1}e^{-i\Phi})\}. \quad (11)$$

Here, $\Phi = nk_0\sin\alpha[x(\cos\phi_2 - \cos\phi_1) + y(\sin\phi_2 - \sin\phi_1)] + \varphi_2 - \varphi_1$ and $t(\alpha) = t_s(\alpha)t_p^*(\alpha) = T(\alpha)e^{i\delta(\alpha)}$ with $T(\alpha)$ being the modulus and $\delta(\alpha)$ the phase of $t(\alpha)$. $T(\alpha)$ and $\delta(\alpha)$ originally come from the transmittance and the transmission phase shift of the *s*- and *p*-pol. component of the incident light in TIR. Once $E_j$, $\phi_{1,2}$, $\Delta\theta_{1,2}$, and $\varphi_{1,2}$ are chosen, the analytical solution of OC can be obtained from eq (11). Again, for a fair comparison, we set the total intensity of the incident light beams to be the same as that of free-space CPL and the OC values are normalized to $C_{\mathrm{CPL}}$. The results of several cases for the cross-propagating ($\phi_1 = 0$, $\phi_2 = \frac{\pi}{2}$) and counter-propagating ($\phi_1 = 0$, $\phi_2 = \pi$) EWs are listed in Table 2.

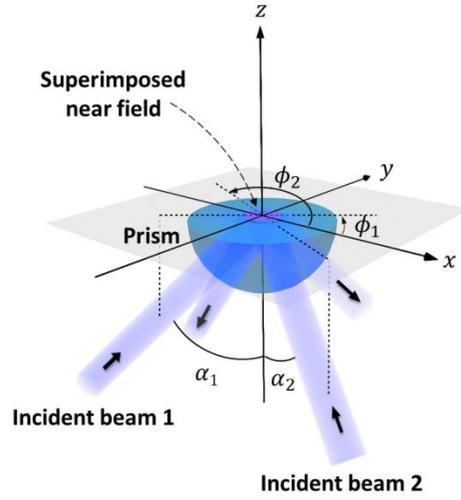

**Figure 3.** Schematic of two incident plane waves totally reflected at the prism-air interface. The incident angle is $\alpha_{1,2}$ and the orientation angle is $\phi_{1,2}$ respect to the $+x$ axis.

| Configuration<br>Incident beams | 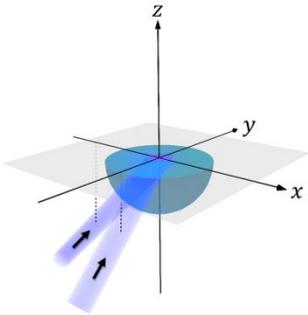<br>Cross-propagation ($\phi_1 = 0, \phi_2 = \frac{\pi}{2}$)<br>$\Phi = -nk_0(x+y)\sin\alpha + \varphi_2 - \varphi_1$<br>$k_C = \sqrt{2}nk_0\sin\alpha$ | 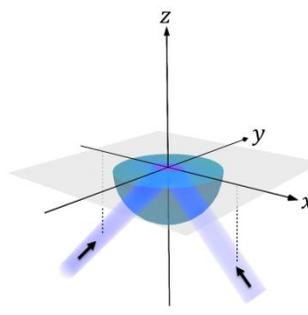<br>Counter-propagation ($\phi_1 = 0, \phi_2 = \pi$)<br>$\Phi = -2nk_0 x\sin\alpha + \varphi_2 - \varphi_1$<br>$k_C = 2nk_0\sin\alpha$ |
|---|---|---|
| $s$- and $p$-pol. light beam | $\hat{C} = -\frac{1}{2}e^{-2\kappa(\alpha)z}n^2\sin^2\alpha T(\alpha)$ $\sin[\Phi - \delta(\alpha)]$ | $\hat{C} = 0$ |
| two $p$- or $s$-pol. light beams | $\hat{C} = 0$ | $\hat{C} = 0$ |
| two CPL beams with same handedness (e.g., left-handed) | $\hat{C} = \frac{1}{2}e^{-2\kappa(\alpha)z}n^2\sin^2\alpha T(\alpha)\cos[\delta(\alpha)]$ $(2+\cos\Phi)$ | $\hat{C} = e^{-2\kappa(\alpha)z}n^2\sin^2\alpha T(\alpha)$ $\cos[\delta(\alpha)]$ |
| two CPL beams with opposite handedness | $\hat{C} = \frac{1}{2}e^{-2\kappa(\alpha)z}n^2\sin^2\alpha T(\alpha)\sin[\delta(\alpha)]$ $\sin\Phi$ | $\hat{C} = 0$ |
| LPL (e.g., $p$-pol.) and CPL (e.g., left-handed) beam | $\hat{C} = \frac{1}{3}e^{-2\kappa(\alpha)z}n^2\sin^2\alpha T(\alpha)$ $\{2\cos[\delta(\alpha)] + \cos[\Phi + \delta(\alpha)]\}$ | $\hat{C} = \frac{2}{3}e^{-2\kappa(\alpha)z}n^2\sin^2\alpha T(\alpha)$ $\cos[\delta(\alpha)]$ |

**Table 2.** Analytical solutions of the normalized OC generated by the superposition of two TIR-based EWs excited by two incident plane waves cross-propagating (second column) and counter-propagating (third column) at incident angles of $\alpha$.

To verify the validity of the analytical solutions, we performed FDTD simulations on several configurations of the incident beams. The theoretically predicted OC patterns are all well reproduced by FDTD simulations (Figs. 4(a)-(j)). Compared to the superposition of two plane waves in free space, the OC patterns generated here are in the near field region bounded to the prism surface. The amplitude of the OC patterns is influenced by the refractive index of the prism $n$ ($n > 1$) and the value of $T(\alpha)$, which is about 6 when $\alpha$ is very close to the critical angle $\alpha_c$ (see Supporting Information, Fig. S.1). Therefore, the near-field OC can be enhanced by applying the TIR-based EWs. For example, the superposition of the EWs of two cross-propagating totally reflected beams with orthogonal polarization state, e.g., $s$-pol. and $p$-pol., yields a pattern with enhanced OC (Fig. 4(a)) compared to the pattern obtained with two far-field cross-polarized beams (Fig. 2(a)). However, no OC can be generated by using other configurations of two incident LPL beams (Figs. 4(b), (f) and (g)). Even higher enhancement in the OC with spatially invariant handedness (Fig. 4(c)) can be generated by the interference of the EWs of two totally reflected CPL beams with the same handedness. Interestingly, the superposition of two cross-propagating EWs excited by two incident CPL beams with the opposite handedness can lead to a non-zero structured OC pattern (Fig. 4(d)). This is very different from the superposition of two far-field CPL beams with the opposite handedness, which results in zero OC (Fig. 2(d)). By analyzing the corresponding solution in Table 2, we found that the non-zero OC comes from the term $\delta(\alpha)$, which is related to the transmission phase shift of the *s*- and *p*-component induced in TIR. By applying two totally reflected CPL beams with the opposite handedness in counter-propagation configuration, the OC of the superimposed EWs remains zero (Fig. 4(i)). This is similar to the case of two counter-propagating far-field CPL beams with the opposite handedness (Fig. 2(i)). Finally, the superposition of the EW of a LPL beam and that of a CPL beam in cross-propagation configuration (Fig. 4(e)) leads to a structured OC pattern with relatively lower OC enhancement compared to Fig. 4(c). If enhanced OC with uniform spatial distribution is needed, one may employ two counter-propagating CPL beams with the same handedness (Fig. 4(h)) or counter-propagating CPL and LPL beam (Fig. 4(j)).

Another important feature of the OC pattern generated by the TIR-based EWs is that, with the same orientation angle difference $|\phi_1 - \phi_2|$, the spatial frequency is $n$ times larger than that in the scenario of far-field plane waves. This can be easily confirmed by identifying the finer OC stripes in Figs. 4(a), (c) and (e) compared to those in Figs. 2(a), (c) and (e). It is a direct consequence of the larger wave vector of EWs compared to that of the plane waves in free space. For trapping and sorting applications, enhanced OC with higher spatial frequency provides larger spatial gradient of the OC and thus a larger reactive component of the chiral optical force.[31] For high resolution chiral domain imaging[43], higher spatial frequency of the OC pattern promises higher spatial resolution.

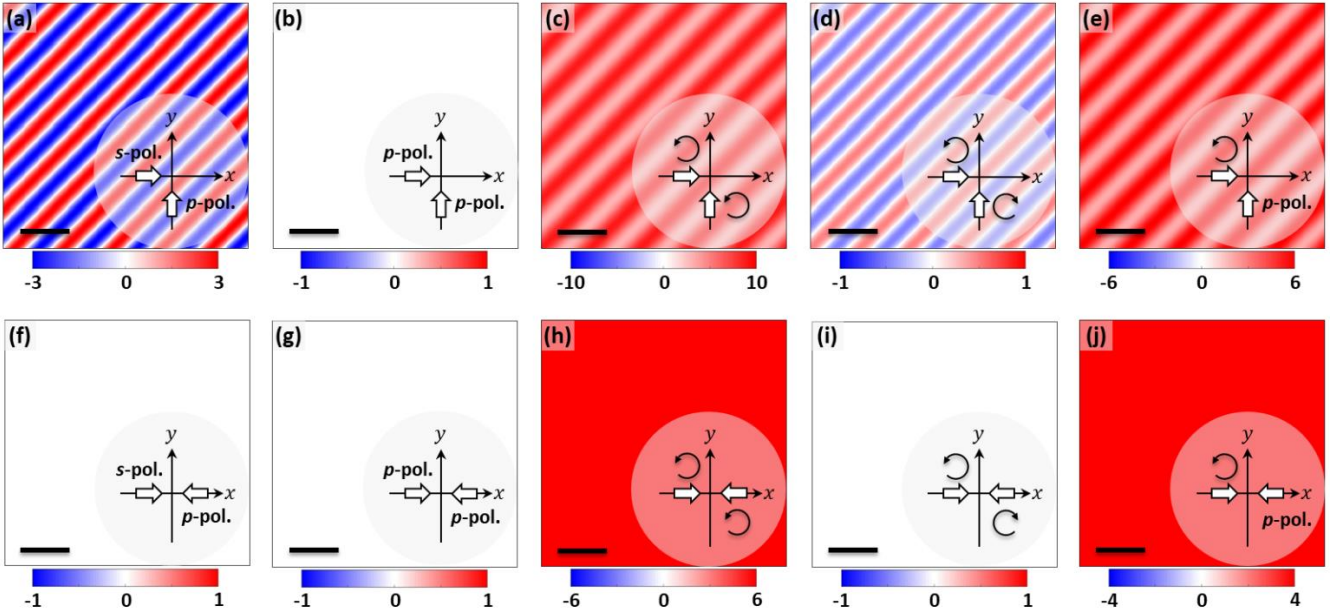

**Figure 4.** Numerically simulated normalized OC patterns generated by the superposition of two TIR-based EWs excited by (a,f) a *s*- and a *p*-pol. light beam; (b,g) two *p*- or *s*-pol. light beams; (c,h) two same-handed CPL beams (e.g., left-handed); (d,i) two opposite-handed CPL beams and (e,j) a *p*-pol. and a left-handed CPL beam. The two incident beams are (a-e) cross-propagating and (f-j) counter-propagating at an incident angle of $\alpha_1 = \alpha_2 = 41°$. The insets indicate the polarization states of the two incident beams and the projection of the incident directions on the *xy* plane. Scale bars: $2\pi/k_0$.

## Superposition of two surface plasmon waves

To further increase the spatial frequency of the OC pattern, one may exploit surface plasmon waves. In this section, we derive the general analytical solution of OC generated by the superposition of two SPWs propagating in arbitrary in-plane (*xy* plane) directions $\phi_{1,2}$ respect to the +*x* axis (Fig. 5(a)). Considering the transverse magnetic nature of the SPWs, the magnetic field of each SPW is described by

$$\widetilde{\mathbf{B}}_{1,2} = B_{SP} e^{iqz} e^{ik_{SP}(x\cos\phi_{1,2}+y\sin\phi_{1,2})+i(-\omega t+\varphi_{1,2})}(-\sin\phi_{1,2}\hat{\mathbf{x}} + \cos\phi_{1,2}\hat{\mathbf{y}}), \quad (12)$$

where $B_{SP}$ is the magnetic field amplitude and $\varphi_{1,2}$ is the initial phase of each SPW. $k_{SP} = k_{SP}' + ik_{SP}''$ and $q = q' + iq''$ are the in-plane and out-of-plane (along *z*-axis) complex wave vectors satisfying the relationship $k_{SP}^2 + q^2 = \left(\frac{\omega}{c}\right)^2$. Accordingly, the electric field of each SPW can be expressed as

$$\widetilde{\mathbf{E}}_{1,2} = -\frac{c^2}{i\omega} B_{SP} e^{iqz} e^{ik_{SP}(x\cos\phi_{1,2}+y\sin\phi_{1,2})+i(-\omega t+\varphi_{1,2})}(-iq\cos\phi_{1,2}\hat{\mathbf{x}} - iq\sin\phi_{1,2}\hat{\mathbf{y}} + ik_{SP}\hat{\mathbf{z}}), \quad (13)$$

and the OC distribution of the superimposed SPWs is

$$C = -\varepsilon_0 q' E_{SP}^2 e^{-2q''z} e^{-k_{SP}''[x(\cos\phi_1+\cos\phi_2)+y(\sin\phi_1+\sin\phi_2)]} \sin(\phi_1 - \phi_2)\sin\Phi, \quad (14)$$

where $\Phi = k_{SP}'[x(\cos\phi_2 - \cos\phi_1) + y(\sin\phi_2 - \sin\phi_1)] + \varphi_2 - \varphi_1$, $E_{SP}$ is the electric field amplitude of each SPW, and the two exponential terms denote the out-of-plane and in-plane damping factors, respectively. Equation (14) indicates that the OC of the superimposed SPWs is proportional to $q'$, the real part of the out-of-plane wave vector of the SPWs. It is worth noting that the out-of-plane wave vector of SPWs possesses a small but non-zero real part (Table S.1, Supporting Information), whereas that of the EWs in TIR is purely imaginary. This is an important difference between these two near fields and is the reason why the superposition of the EWs excited by two totally reflected *p*-pol. light beams is achiral (Figs. 4(b) and (g)). Equation (14) also suggests that a near-field OC pattern can be generated as long as the propagating directions of the two SPWs are not parallel to each other, i.e., $\sin(\phi_1 - \phi_2) \neq 0$.

To verify the analytical prediction, the near-field OC patterns are numerically simulated using the FDTD method. Mode sources have been used to launch the SPWs. The orientation angles of the SPWs are set to $\phi_1 = 0$, $\phi_2 = \frac{\pi}{2}$ for cross-propagation and $\phi_1 = 0$, $\phi_2 = \pi$ for counter-propagation configuration (see Methods for details). The superposition of two propagating SPWs results in non-zero OC (Fig. 5(b)), except for the case of counter-propagation configuration (Fig. 5(c)). The magnitude of the OC pattern in Fig. 5(b) decreases along the propagating directions of the SPWs because of the plasmon damping. To quantify the OC enhancement relative to free-space CPL, a factor $m$ which describes the field intensity enhancement between the excitation light and the excited SPWs should be considered. This factor depends on the methods used for photon-to-plasmon coupling, e.g., Kretschmann configuration at the best coupling angle yields $m \sim 10^2$.[48] According to the simulated OC values normalized to $C_{CPL}$ (Fig. 5(b)), an OC enhancement larger than 1 requires $m$ to be at least 50. From eq (14), the amplitude of the OC on the metal surface ($z \sim 0$) depends on the real part of the out-of-plane wave vector $q'$ and the near-field intensity. For SPWs, the quantity $q'$ is about two orders of magnitude smaller than the wavenumber of the light in vacuum $k_0$ (Table S.1, Supporting Information). Therefore, within the interference area, the enhancement of OC is always smaller than that of the near-field intensity. Nevertheless, the spatial frequency of the structured OC pattern generated here is higher than the corresponding cases in the previous two scenarios owing to the relatively large wave vector of the SPWs.

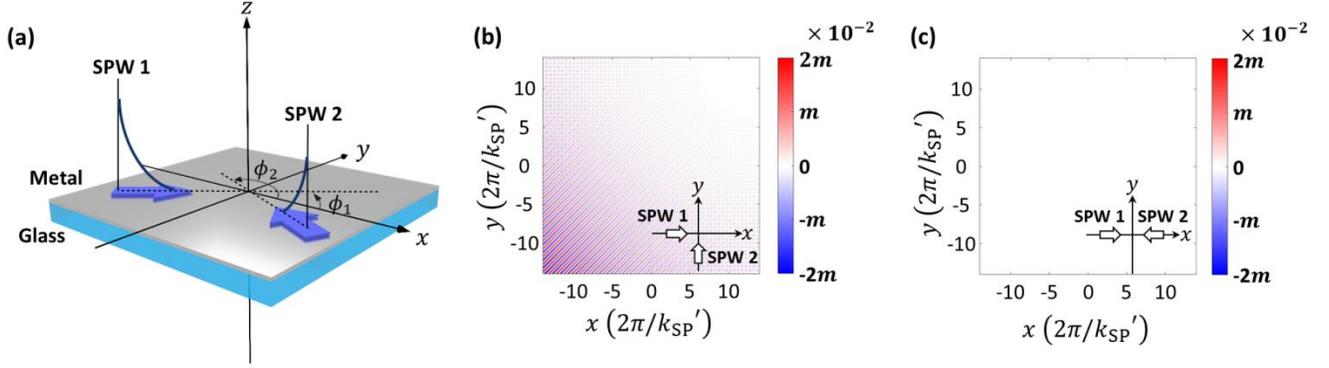

**Figure 5.** Generation of OC patterns by the superposition of two SPWs. (a) Schematic of two SPWs propagating in arbitrary in-plane directions $\phi_{1,2}$ respect to the $+x$ axis. Numerically simulated normalized OC patterns generated by (b) two cross-propagating SPWs and (c) two counter-propagating SPWs. The insets in (b) and (c) show the top views of the SPWs.

## Conclusion

We have theoretically investigated the generation of OC patterns by the superposition of two waves in three scenarios, namely free-space plane waves, TIR-based EWs and propagating SPWs. In each scenario, the general analytical solution of OC is derived and several practically realizable configurations are discussed. The analytically predicted OC patterns are all well reproduced by FDTD numerical simulations. The results show that spatially structured OC patterns can be generated with far-field waves as well as near-field EWs and SPWs. Enhancement in OC can be obtained from the constructive interference of free-space CPL, the enhanced EWs in TIR, and the SPWs with sufficiently high intensity enhancement. This work may serve as a guideline for generating OC patterns by two-wave-superposition for applications in chirality sorting, chiral imaging and snapshot CD measurement.

## Methods

**Numerical Simulations.** To verify the analytical predictions, three-dimensional full-wave FDTD method (FDTD Solutions, Lumerical) has been employed to simulate the OC patterns. In all of the simulations, optical responses at 405 nm were recorded. The left-handed CPL was simulated by superposing two plane waves at orthogonal polarization state with a phase difference of $\frac{\pi}{2}$. The electric field amplitude of each plane wave was set to 1 V/m and the simulated magnitude of OC was $1.3 \times 10^{-4}$ J/m$^4$. In the first two scenarios, plane wave sources were used. The incident angles of the plane waves and the incident light beams in TIR $\alpha$ were set to $41°$. In the scenario of SPWs, mode sources were used to inject the propagating SPWs. The amplitude of each mode source was set to 1 V/m, and the simulated OC was multiplied by the intensity enhancement factor $m$. The refractive index of the glass substrate in the scenarios of TIR-based EWs and SPWs has been set to 1.5396. For the scenario of SPWs, silver with permittivity $\varepsilon_{Ag} = -4.7149 + i0.2170$ at 405 nm has been used as the material of the metallic film. For all simulations, a 2D *xy* plane monitor was employed to record the electromagnetic field. To measure the near field, the monitor was positioned 5 nm above the glass and metal surface. In the first two scenarios, the simulation region was set as 2 μm (*x*) × 2 μm (*y*) × 2 μm (*z*) with Bloch boundaries in *x* and *y* directions and perfectly matched layer (PML) boundaries in *z* directions. Uniform spatial discretization of 5 nm has been applied to *x*, *y* and *z* directions within a mesh override region of 2 μm (*x*) × 2 μm (*y*) × 0.4 μm (*z*) at the center of the simulation box. In the scenario of SPWs, the simulation region was set to 10 μm (*x*) × 10 μm (*y*) × 2 μm (*z*) with metal boundaries in the propagating direction of the SPW, periodic boundaries in the orthogonal direction of the propagation direction on the *xy* plane and PML boundaries in *z* directions. Here, a large *xy* simulation plane was used to visualize the propagation damping of the SPWs. Spatial discretization of 25 nm in *x*, *y* directions and 5 nm in *z* direction has been applied within a mesh override region of 10 μm (*x*) × 10 μm (*y*) × 0.4 μm (*z*) at the center of the simulation box.

## Acknowledgement

The support from the DFG (HU2626/3-1 and CRC 1375 NOA) is acknowledged. J.Z. acknowledges the support from Sino-German (CSC-DAAD) Postdoc Scholarship Program, 2018.## Supporting Information

Supporting Information Available: <Transmittance of the incident light in total internal reflection, Calculation parameters for the surface plasmon waves.> This material is available free of charge via the Internet at http://pubs.acs.org

# Reference

1. Lipkin, D. M. Existence of a new conservation law in electromagnetic theory. *J. Math. Phys.* **1964**, *5*, 696−700.
2. Tang, Y.; Cohen, A. E. Optical chirality and its interaction with matter. *Phys. Rev. Lett.* **2010**, *104*, 163901.
3. Tang, Y.; Cohen, A. E. Enhanced enantioselectivity in excitation of chiral molecules by superchiral light. *Science* **2011**, *332*, 333−336.
4. Hendry, E.; Carpy, T.; Johnston, J.; Popland, M.; Mikhaylovskiy, R. V.; Lapthorn, A. J.; Kelly, S. M.; Barron, L. D.; Gadegaard, N.; Kadodwala, M. Ultrasensitive detection and characterization of biomolecules using superchiral fields. *Nat. Nanotechnol.* **2010**, *5*, 783−787.
5. Hu, H.; Gan, Q.; Zhan, Q. Generation of a nondiffracting superchiral optical needle for circular dichroism imaging of sparse subdiffraction objects. *Phys. Rev. Lett.* **2019**, *122*, 223901.
6. Schäferling, M.; Yin, X.; Giessen, H. Formation of chiral fields in a symmetric environment. *Opt. Express* **2012**, *20*, 26326−26336.
7. Eftekhari, F.; Davis, T. J. Strong chiral optical response from planar arrays of subwavelength metallic structures supporting surface plasmon resonances. *Phys. Rev. B: Condens. Matter Mater. Phys.* **2012**, *86*, 075428.
8. Schäferling, M.; Dregely, D.; Hentschel, M.; Giessen, H. Tailoring enhanced optical chirality: design principles for chiral plasmonic nanostructures. *Phys. Rev. X* **2012**, *2*, 031010.
9. Valev, V. K.; Baumberg, J. J.; Sibilia, C.; Verbiest, T. Chirality and chiral optical effects in plasmonic nanostructures: fundamentals, recent progress, and outlook. *Adv. Mater.* **2013**, *25*, 2517−2534.
10. Davis, T. J.; Hendry, E. Superchiral electromagnetic fields created by surface plasmons in nonchiral metallic nanostructures. *Phys. Rev. B: Condens. Matter Mater. Phys.* **2013**, *87*, 085405.
11. Schäferling, M.; Yin, X.; Engheta, N.; Giessen, H. Helical plasmonic nanostructures as prototypical chiral near-Field sources. *ACS Photonics* **2014**, *1*, 530−537.
12. Hashiyada, S.; Narushima, T.; Okamoto, H. Local optical activity in achiral two-dimensional gold nanostructures. *J. Phys. Chem. C* **2014**, *118*, 22229−22233.
13. Lin, D.; Huang, J.-S. Slant-gap plasmonic nanoantennas for optical chirality engineering and circular dichroism enhancement. *Opt. Express* **2014**, *22*, 7434−7445.
14. Tian, X.; Fang, Y.; Sun, M. Formation of enhanced uniform chiral fields in symmetric dimer nanostructures. *Sci. Rep.* **2015**, *5*, 17534.
15. Nesterov, M. L.; Yin, X.; Schäferling, M.; Giessen, H.; Weiss, T. The role of plasmon-generated near fields for enhanced circular dichroism spectroscopy. *ACS Photonics* **2016**, *3*, 578−583.
16. Schäferling, M.; Engheta, N.; Giessen, H.; Weiss, T. Reducing the complexity: enantioselective chiral near-fields by diagonal slit and mirror configuration. *ACS Photonics* **2016**, *3*, 1076−1084.
17. Zu, S.; Bao, Y.; Fang, Z. Planar plasmonic chiral nanostructures. *Nanoscale*, **2016**, *8*, 3900−3905.
18. Kramer, C.; Schäferling, M.; Weiss, T.; Giessen, H.; Brixner, T. Analytic optimization of near-field optical chirality enhancement. *ACS photonics* **2017**, *4*, 396−406.
19. Luo, Y.; Chi, C.; Jiang, M.; Li, R.; Zu, S.; Li, Y.; Zheyu, F. Plasmonic chiral nanostructures: chiral optical effects and applications. *Adv. Opt. Mater.* **2017**, *5*, 170040.
20. Collins, J. T.; Kuppe, C.; Hooper, D. C.; Sibilia, C.; Centini, M.; Valev, V. K. Chirality and chiral optical effects in metal nanostructures: fundamentals and current trends. *Adv. Opt. Mater.* **2017**, *5*, 1700182.
21. Hentschel, M.; Schäferling, M.; Duan, X.; Giessen, H.; Liu, N. Chiral plasmonics. *Sci. Adv.* **2017**, *3*, e1602735.
22. García-Guirado, J.; Svedendahl, M.; Puigdollers, J.; Quidant, R. Enantiomer-selective molecular sensing using racemic nanoplasmonic arrays. *Nano Lett.* **2018**, *18*, 6279−6285.
23. Tseng, M. L.; Lin, Z.-H.; Kuo, H. Y.; Huang, T.-T.; Huang, Y.-T., Chung, T. L.; Chu, C. H.; Huang, J.-S.; Tsai, D. P. Stress-induced 3D chiral fractal metasurface for enhanced and stabilized broadband near-field optical chirality. *Adv. Opt. Mater.* **2019**, *7*, 1900617.

24. Wang, Z.; Teh, B. H.; Wang, Y.; Adamo, G.; Teng, J.; Sun, H. Enhancing circular dichroism by super chiral hot spots from a chiral metasurface with apexes. *Appl. Phys. Lett.* **2017**, *110*, 221108.
25. Mohammadi, E.; Tsakmakidis, K. L.; Askarpour, A. N.; Dehkhoda, P.; Tavakoli, A.; Altug, H. Nanophotonic platforms for enhanced chiral sensing. *ACS Photonics* **2018**, *5*, 2669−2675.
26. Solomon, M. L.; Hu, J.; Lawrence, M.; García-Etxarri, A.; Dionne, J. A. Enantiospecific optical enhancement of chiral sensing and separation with dielectric metasurfaces. *ACS Photonics* **2019**, *6*, 43−49.
27. Yao, K.; Zheng, Y. Near-ultraviolet dielectric metasurfaces: from surface-enhanced circular dichroism spectroscopy to polarization-preserving mirrors. *J. Phys. Chem. C* **2019**, *123*, 11814−11822.
28. Zhao, X.; Reinhard, B. M. Switchable chiroptical hot-spots in silicon nanodisk dimers. *ACS Photonics* **2019**, *6,* 1981−1989.
29. Pellegrini, G.; Finazzi, M.; Celebrano, M.; Duò, L.; Biagioni, P. Chiral surface waves for enhanced circular dichroism. *Phys. Rev. B: Condens. Matter Mater. Phys.* **2017**, *95*, 241402(R).
30. Pellegrini, G.; Finazzi, M.; Celebrano, M.; Duò, L.; Biagioni, P. Surface-enhanced chiral optical spectroscopy with superchiral surface waves. *Chirality* **2018**, *30*, 883−889.
31. Antoine, C.-D.; James, A. H.; Cyriaque, G.; Thomas, W. E. Mechanical separation of chiral dipoles by chiral light. *New J. Phys.* **2013**, *15*, 123037.
32. Wang, S. B.; Chan, C. T. Lateral optical force on chiral particles near a surface. *Nat. Commun.* **2014**, *5*, 3307.
33. Tkachenko, G.; Brasselet, E. Optofluidic sorting of material chirality by chiral light. *Nat. Commun.* **2014**, *5*, 3577.
34. Canaguier-Durand, A.; Genet, C. Chiral near fields generated from plasmonic optical lattices. *Phys. Rev. A: At., Mol., Opt. Phys.* **2014**, *90*, 023842.
35. Alizadeh, M. H.; Reinhard, B. M. Transverse chiral optical forces by chiral surface plasmon polaritons. *ACS Photonics* **2015**, *2*, 1780−1788.
36. Amaury Hayat; J. P. Balthasar Müller; Capasso, F. Lateral chirality-sorting optical forces. *Proc. Natl. Acad. Sci. U.S.A.* **2015**, *112*, 13190−13194.
37. Chen, H.; Liang, C.; Liu, S.; Lin, Z. Chirality sorting using two-wave-superposition–induced lateral optical force. *Phys. Rev. A: At., Mol., Opt. Phys.* **2016**, *93*, 053833.
38. Rukhlenko, I. D.; Tepliakov, N. V.; Baimuratov, A. S.; Andronaki, S. A.; Gun'ko, Y. K.; Baranov, A. V.; Fedorov, A. V. Completely chiral optical force for enantioseparation. *Sci. Rep.* **2016**, *6*, 36884.
39. Zhang, T.; Mahdy, M. R. C.; Liu, Y.; Teng, J. H.; Lim, C. T.; Wang, Z.; Qiu, C.-W. All-optical chirality-sensitive sorting via reversible lateral forces in superposition fields. *ACS Nano* **2017**, *11*, 4292−4300.
40. Zhang, Q.; Li, J.; Liu, X. Optical lateral forces and torques induced by chiral surface-plasmon-polaritons and their potential applications in recognition and separation of chiral enantiomers. *Phys.Chem.Chem.Phys.* **2019**, *21*, 1308−1314.
41. Li, M.; Yan, S.; Zhang, Y.; Liang Y.; Zhang, P.; Yao, B. Optical sorting of small chiral particles by tightly focused vector beams. *Phys. Rev. A: At., Mol., Opt. Phys.* **2019**, *99*, 033825.
42. Arteaga, O.; El-Hachemi, Z.; Ossikovski, R. Snapshot circular dichroism measurements. *Opt. Express* **2019**, *27*, 6746−6756.
43. Huang, S.-Y.; Zhang, J.; Karras, C.; Förster, R.; Heintzmann, R.; Huang, J.-S. Chiral structured illumination microscopy. *ArXiv:1908.09391*, **2019**.
44. Kruining, K. C.; Cameron, R. P.; Götte, J. B. Superpositions of up to six plane waves without electric-field superposition. *Optica* **2018**, *5*, 1091–1098.
45. Cameron, R. P.; Barnett, S. M.; Yao, A. M. Optical helicity of interfering waves. *J. Mod. Opt.* **2014**, *61*, 25–31.
46. Ohdaira, Y.; Inoue, T.; Hori, H.; Kitahara, K. Local circular polarization observed in surface vortices of optical near-fields. *Opt. Express* **2008**. *16*, 2915–2921.
47. Born, M.; Wolf, E. *Principles of optics*, 7th ed., University Press: Cambridge, 1999.
48. Raether H. *Surface plasmons*, Springer-Verlag, Berlin, 1988.

# – Supporting Information –

# Generation of Optical Chirality Patterns with Plane Waves, Evanescent Waves and Surface Plasmon Waves

Jiwei Zhang,[1,*] Shiang-Yu Huang,[1] Zhan-Hong Lin[1] and Jer-Shing Huang[1,2,3,4,*]

[1]*Leibniz Institute of Photonic Technology, Albert-Einstein Straße 9, 07745 Jena, Germany*
[2]*Abbe Center of Photonics, Friedrich-Schiller University Jena, Jena, Germany*
[3]*Research Center for Applied Sciences, Academia Sinica, 128 Sec. 2, Academia Road, Nankang District, 11529 Taipei, Taiwan*
[4]*Department of Electrophysics, National Chiao Tung University, 1001 University Road, 30010 Hsinchu, Taiwan*

*e-mail address: jiwei.zhang@leibniz-ipht.de; jer-shing.huang@leibniz-ipht.de*

## S.1 Transmittance of the incident light in total internal reflection

The confined evanescent wave in total internal reflection possesses an enhanced electric field compared to the incident light beam. The enhancement is determined by the transmission coefficient calculated by Fresnel equations. The relationship between the transmittance and the incident angle is plotted in Fig. S.1. The calculation parameters are consistent with those in the numerical simulations.

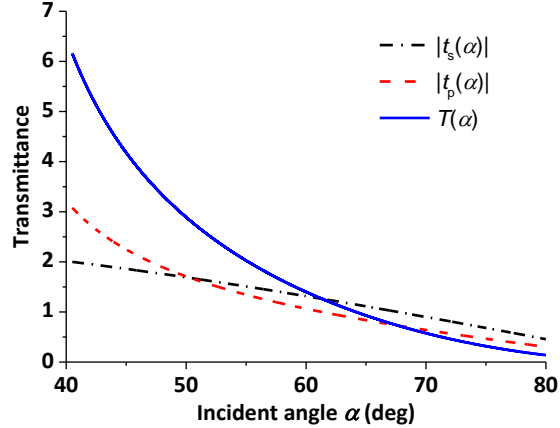

**Figure S.1.** Transmittance of the *s*- ($|t_s(\alpha)|$) and *p*-polarized component ($|t_p(\alpha)|$) of the incident light, and $T(\alpha)$ change with the incident angle $\alpha$.

## S.2 Calculation parameters for the surface plasmon wave

Table S.1 summarizes the data of a surface plasmon wave (SPW) excited at 405 nm. The real part of the out-of-plane wave vector $q'$ ($2.3454\times10^5$ m$^{-1}$) is about 100 times smaller than $k_0$ ( $= 1.5514\times10^7$ m$^{-1}$).

| | |
|---|---|
| $\lambda$ | 405 nm |
| $k_0$ | $1.5514 \times 10^7$ m$^{-1}$ |
| $\varepsilon_{Ag}$ | $-4.7149 + i0.2170$ |
| $\varepsilon_{air}$ | 1 |
| $k_{SP} = \dfrac{2\pi}{\lambda}\sqrt{\dfrac{\varepsilon_{Ag}\varepsilon_{air}}{\varepsilon_{Ag}+\varepsilon_{air}}}$ | $1.7472 \times 10^7 + i1.0791 \times 10^5$ m$^{-1}$ |
| $q = \sqrt{k_0{}^2 - k_{SP}{}^2}$ | $2.3454 \times 10^5 - i8.0389 \times 10^6$ m$^{-1}$ |

**Table S.1.** Calculation parameters for the SPW.